\documentclass[a4paper,11pt]{article}
\usepackage{jheppub} 
\usepackage{lineno}
\usepackage{verbatim}
\usepackage{multirow}
\usepackage{booktabs}
\usepackage{subcaption}

\title{\boldmath Flattenicity dependence of charge and transverse-momentum balance functions in $pp$ collisions at $\sqrt{s}=13.6$ TeV with PYTHIA8}

\author[a]{Srabani Dutta Roy,}
\author[a]{Rohit Agarwala,}
\author[a,1]{Kalyan Dey\note{Corresponding author.}}
\affiliation[a]{Department of Physics, Bodoland University, Kokrajhar-783370, Assam, India}

\emailAdd{kalyan.dey@buniv.edu.in}

\abstract{We study the charge balance function $B(\Delta\eta,\Delta\phi)$ and the charge-dependent transverse-momentum balance function
$P_{2}^{\rm CD}(\Delta\eta,\Delta\phi)$ in proton--proton ($pp$) collisions at $\sqrt{s}=13.6$~TeV using \texttt{PYTHIA~8.315} with the Monash tune. Flattenicity is employed as an event-shape observable to probe how these correlations depend on the topology of particle production. The results are examined across different charged-particle multiplicity intervals to disentangle the effects of overall event activity and event topology on $B$ and $P_{2}^{\rm CD}$.
Both observables exhibit pronounced near-side correlations around
$(\Delta\eta,\Delta\phi)\approx(0,0)$. The near-side peaks become stronger with increasing charged-particle multiplicity, and the RMS widths of both observables decrease systematically, indicating that the correlations become more localized in relative pseudorapidity and azimuth. A clear dependence on flattenicity is also observed: jet-like events (low flattenicity) generally exhibit stronger and narrower near-side correlations than isotropic events (high flattenicity). This
event-shape dependence is more pronounced for $P_{2}^{\rm CD}$ than for $B$, indicating that $P_{2}^{\rm CD}$ is more sensitive to event topology, whereas $B$ primarily probes charge-dependent particle-pair correlations. These results demonstrate the complementary nature of the two observables and the potential of flattenicity-based event selection for studying particle-production mechanisms in high-multiplicity $pp$ collisions.}

\begin{document}
\maketitle
\flushbottom

\section{Introduction}
\label{introduction}
Smashing heavy ions at ultra-relativistic energy is the only avenue to understand the properties and dynamics of strongly interacting matter under extreme conditions. Experiments at the Relativistic Heavy Ion Collider (RHIC) at Brookhaven
National Laboratory provided compelling evidence for the formation of a strongly interacting, deconfined state of quarks and gluons, characterized by strong collective behaviour
\cite{STAR:2005gfr,Christiansen:2024bhe}. Subsequent measurements at CERN's Large Hadron Collider (LHC) have further advanced our understanding of the properties and dynamical evolution of the quark--gluon plasma (QGP) \cite{Braun-Munzinger:2007edi}. Interestingly, several features traditionally associated with heavy-ion collisions have also been observed in high-multiplicity proton--proton ($pp$) collisions at the LHC, including long-range
angular correlations \cite{Dumitru:2010iy}, enhanced strange-hadron production \cite{ALICE:2016fzo}, and collective-like behaviour indicated by the number-of-constituent-quark (NCQ) scaling of the elliptic flow coefficient ($v_2$) \cite{ALICE:2024vzv}. These observations have raised important questions about the origin of correlations in small collision systems, particularly whether they arise from collective dynamics or from microscopic particle-production
mechanisms. Consequently, $pp$ collisions provide an important
opportunity to investigate particle-production dynamics beyond their traditional role as a reference for heavy-ion measurements.

 Single-particle observables, such as charged-particle multiplicity, transverse-momentum spectra, and pseudorapidity distributions, provide information regarding the overall characteristics of particle production. However, these observables do not fully characterize correlations arising from local conservation laws and the dynamics of particle production. Two-particle correlation observables provide a more differential probe by retaining information on the relative kinematics of particle pairs and can therefore provide additional insight into the mechanisms governing particle production and hadronisation.

The charge balance function is one of the most powerful probes of local charge conservation. Since in every microscopic particle production process, electric charge is conserved, the oppositely charged particles are produced in balancing pairs that maintain their correlation throughout the evolution of the medium. The relative separation of these balancing partners in phase space $(\eta, \phi)$ space is governed by several physical processes such as time of hadronization, parton fragmentation, resonance decays, diffusion, collectivity and final--state interactions \cite{Pratt:2021xvg}. Later--produced balancing charges have a narrow correlation structure due to their proximity in phase space, while earlier--produced balancing pairs have a wider spread in their correlation structure in relative pseudorapidity $(\Delta \eta)$ and azimuthal angle $(\Delta \phi)$. Consequently, these balancing charge pair correlations provide insight into the space--time evolution of particle production and hadronization mechanisms \cite{Pratt:2018ebf,Pratt:2021xvg}.

To quantify charge correlations, Bass, Danielewicz, and Pratt proposed the charge balance function as a statistical measure of the conditional probability of finding a balancing charge associated with a given charged particle \cite{Bass:2000az,Trainor:2003cn}. By combining correlations between oppositely and like-charged particles, the balance function isolates the charge-dependent component associated with local charge conservation and provides sensitivity to the mechanisms and timescales of charge production. It has subsequently been extensively studied by the STAR \cite{STAR:2010plm}, ALICE \cite{ALICE:2013vrb}, and CMS \cite{CMS:2023sua} collaborations across different collision systems and energies.

These experimental measurements have shown that the balance function is strongly dependent on the collision system, collision centrality, and charged-particle multiplicity \cite{ALICE:2015nuz,ALICE:2013tla}. Heavy-ion collisions show a progressive narrowing of the balance function with increasing collision centrality in both $\Delta\eta$ and $\Delta\phi$ \cite{ALICE:2015nuz,ALICE:2013tla,Pan:2018dsq}. This phenomenon has been attributed to a combination of delayed charge production, reduced diffusion of balancing charges, and the formation of strong collective radial expansion during the evolution of the medium \cite{ALICE:2015nuz,Pan:2018dsq}. In proton-proton and proton-nucleus collisions, the balance function shows a clear multiplicity dependence \cite{ALICE:2013tla, Pan:2018dsq, Sicking:2012up}, indicating that microscopic particle production mechanisms have a significant influence on the correlation strength and spatial separation of balancing charge pairs.

However, interpreting the balance-function width is difficult since the observed correlations are caused by numerous competing physical mechanisms. Apart from the impacts of charge creation time and diffusion, resonance decays, parton fragmentation, and hadronization modify the relative separation of balancing charged particles and consequently influence the shape and width of the balance function \cite{Morsch:2014bka,Sarma:2021qfr}. In heavy-ion collision systems, collective radial flow can further narrow the balance function by providing a common transverse boost to balancing charged pairs. In contrast, proton-proton collisions are likely to be dominated by non-perturbative QCD processes, with multi-parton interactions (MPI), color reconnection (CR), and string fragmentation all playing important roles in shaping the observed correlation structure \cite{ALICE:2015nuz}. As a result, measurements of the charge balance function continue to provide important insights into the interplay of conservation principles and microscopic dynamics that govern particle production in a range of collision systems.

While the charge balance function provides information on correlations associated with local charge conservation, momentum conservation provides another important constraint on particle production in high-energy collisions. In this context, the charge-dependent transverse momentum correlator, $P_2^{CD}$, provides a complementary observable for studying transverse-momentum correlations between charged-particle pairs \cite{STAR:2005vxr,Basu:2020ldt}. Unlike the charge balance function, which characterizes the relative distribution of balancing electric charges, $P_2^{CD}$ quantifies charge-dependent correlations in the transverse momenta of particle pairs. It thus offers additional insight into the mechanisms governing particle production and hadronisation \cite{STAR:2005vxr,Basu:2020ldt,Sharma:2008qr,ALICE:2024cms}.

Various physical processes, such as collective expansion, jet fragmentation, resonance decays, and multi-parton interactions, have been found to influence transverse momentum correlations in $p+p$, $p+A$ and $A+A$ collisions \cite{STAR:2005vxr, Basu:2020ldt,Sharma:2008qr,ALICE:2018jco} through both experimental and theoretical investigations. Because charge and momentum conservation are based on different physical principles, studying both the charge balance function and the transverse momentum balance function provides a more comprehensive understanding of the underlying dynamics of particle production than studying either observable separately. Such combined study can help elucidate the
interplay between conservation laws, event-by-event fluctuations,
and the microscopic mechanisms governing particle production in
proton--proton collisions.

Event-shape observables have emerged as effective tools for defining the topology of particle production in high-energy collisions. Observables like transverse sphericity \cite{ALICE:2012cor}, transverse spherocity \cite{Mallick:2020dzv,ALICE:2019dfi,Bencedi:2018ctm}, and, more recently, flattenicity \cite{Ortiz:2022mfv,Ortiz:2022zqr} can classify collision events based on their geometrical properties, providing insight into the relative contributions of hard and soft QCD processes. Transverse spherocity, in particular, has been widely employed to distinguish between jet-like and isotropic events using the azimuthal distribution of transverse momentum of charged particles produced in the collision. Flattenicity quantifies uniformity of particle production over the two-dimensional $\eta$--$\phi$ phase space, providing a more comprehensive characterisation of event topology. Thus, unlike spherocity, which is primarily based on the transverse momentum distribution in pseudorapisity, flattenicity takes into account both pseudorapidity and azimuthal information, allowing for a more complete description of the spatial distribution of particle production.

Event-shape observables have been used to investigate the dependence of balance functions on the topology of particle production in proton--proton ($pp$) collisions. Studies employing transverse spherocity have reported differences in balance-function correlations between jet-like and isotropic events, with jet-like events exhibiting narrower correlation peaks than isotropic events \cite{Behera:2025ycg,Behera:2025ymi}. This behaviour may be associated with the more localized particle production such as, jet fragmentation in jet-like events, whereas isotropic events can exhibit broader correlations possibly due to enhanced contribution of multi parton interactions (MPI) \cite{Behera:2025ycg,Behera:2025ymi}. These findings demonstrate the relevance of event topology when studying charge-dependent particle correlations. However, the dependence of the charge balance function and the charge-dependent transverse-momentum balance function on flattenicity remains largely unexplored. Investigating the two-particle charge balance function and the charge-dependent transverse momentum balance function as a function of flattenicity therefore provides a complementary perspective on the interplay between charge conservation, momentum conservation, and the spatial topology of particle production in proton--proton collisions. In particular, the use of flattenicity allows these correlations to be studied with an event-shape classifier that simultaneously accounts for particle-density fluctuations in both $\eta$ and $\phi$.

In this work, we study charge and transverse-momentum balance functions in pp collisions at $\sqrt{s}= 13.6$ TeV using \texttt{PYTHIA8} \cite{Sjostrand:2014zea}, and investigate their dependence on event topology through flattenicity. The remainder of this paper is organized as follows. Section~\ref{Methodology} describes the \texttt{PYTHIA8} model simulations, flattenicity estimation
classification procedure and the methodology used to calculate the balance functions and their widths. Section~\ref{results} presents and discusses the results for the charge balance function and the charge-dependent transverse-momentum balance function as a function of charged-particle multiplicity and flattenicity. Finally, section~\ref{summary} summarizes the main findings of this study and presents the conclusions.

\section{Methodology}
\label{Methodology}
\subsection{Event Generation and Classification}
\texttt{PYTHIA8} \cite{Sjostrand:2014zea, Sjostrand:2006za, Sjostrand:2007gs, Bierlich:2022pky} is a general--purpose Monte Carlo event generator commonly used to simulate ultra--relativistic collisions between leptons, protons and nuclei. It includes QCD calculations for hard partonic scatterings as well as a variety of physical processes like initial and final state parton showers, multi parton interactions (MPI), color reconnection (CR) and string fragmentation. Although the MPI property of \texttt{PYTHIA8} accurately describes the experimental data for small systems \cite{D0:2009msb, ZEUS:2007idj} it fails to accurately predict the transverse momentum distribution \cite{ALICE:2015juo}, multiplicity dependence of strange particle production \cite{ALICE:2011gmo} and jet production rate \cite{ALICE:2021clt} at LHC energies. The \texttt{PYTHIA8} model therefore, is incorporated with a new final state phenomenon called color recnnection (CR) mechanism \cite{Argyropoulos:2014pka,Goswami:2019mta}. Colour reconnection (CR) is a hadronization-related mechanism in which colour connections between partons are rearranged before hadronization, thereby influencing the final-state particle distributions and correlations \cite{Argyropoulos:2014pka,Goswami:2019mta}

Implementing CR mechanism in \texttt{PYTHIA8} model provides an explanation for several other experimental phenomena like multiplicity dependence of mean transverse momentum \cite{ALICE:2013ccj}, and the linear rise in $J/\psi$ with multiplicity at forward rapidities \cite{ALICE:2012bny, Thakur:2017msb} in $p+p$ collisions. The CR process therefore reflects several heavy-ion like effects in $p+p$ collisions.

The present analysis utilises a total of approximately 388 million minimum-bias $p+p$ collision events generated using \texttt{PYTHIA 8.315} with the Monash 2013 tune at a center-of-mass energy of $\sqrt{s}=13.6$ TeV. The parameter settings utilized in the simulation of the $pp$ collision events are tabulated in table \ref{tab:simulation}.

\begin{table}[htbp]
\centering

\begin{tabular}{|l l|}
\hline
\textbf{Parameter} & \textbf{Value}\\
\hline
Collision system & $pp$ \\
Centre-of-mass energy & $\sqrt{s}=13.6$~TeV \\
Event generator & \texttt{PYTHIA~8.315} \\
Tune & Monash 2013 (Tune:pp = 14) \\
Physics process & SoftQCD Non-Diffractive \\
Colour reconnection & MPI-based ($RR=1.8$) \\
Generated events & $\sim388$ million \\
\hline
\end{tabular}
\caption{PYTHIA~8.315 simulation settings used in the present analysis.\label{tab:simulation}}
\end{table}

Charged particles satisfying $|\eta|<0.8$ and
$0.2<p_{\rm T}<5.0$~GeV/$c$ are used to determine event multiplicity
and calculate the charge balance function, $B$, and the
charge-dependent transverse-momentum balance function,
$P_{2}^{\rm CD}$, defined in section~\ref{B} and ~\ref{P2} respectively. Events are divided into four charged-particle
multiplicity classes and further classified according to flattenicity
within each multiplicity class, as described in
section~\ref{flattenicity}. The corresponding selection criteria
and analysis ranges are summarized in Table~\ref{tab:selection},
while the corresponding average charged-particle multiplicities for the event classes are given in Table~\ref{tab:eventclass}.



\begin{table}[htbp]
\centering

\begin{tabular}{|l l|}
\hline
\textbf{Selection criterion} & \textbf{Requirement} \\
\hline
Pseudorapidity acceptance & $|\eta|<0.8$ \\
Transverse-momentum range & $0.2<p_{\rm T}<5.0$~GeV/$c$ \\
$\Delta\eta$ range & $[-1.6,1.6]$ \\
$\Delta\phi$ range & $[-\pi/2,3\pi/2]$ \\
Width calculation range in $\Delta\eta$ & $|\Delta\eta|\leq 1.0$ \\
Width calculation range in $\Delta\phi$ & $|\Delta\phi|\leq \pi/2$ \\
\hline
\end{tabular}
\caption{Particle selection criteria and correlation ranges used in the analysis.}
\label{tab:selection}
\end{table}

\begin{table}[htbp]
\centering

\begin{tabular}{|c l l|}
\hline
\textbf{Class} & \textbf{Event selection} & \textbf{$\langle N_{\rm ch}\rangle$}\\
\hline
\multirow{3}{*}{I}
& Integrated & $33.40\pm0.002$ \\
& $1-\rho$ $(0$--$20\%)$ & $30.54\pm0.003$ \\
& $1-\rho$ $(80$--$100\%)$ & $36.27\pm0.003$ \\


\multirow{3}{*}{II}
& Integrated & $21.65\pm0.001$ \\
& $1-\rho$ $(0$--$20\%)$ & $21.16\pm0.001$ \\
& $1-\rho$ $(80$--$100\%)$ & $22.15\pm0.001$ \\


\multirow{3}{*}{III}
& Integrated & $14.65\pm0.0003$ \\
& $1-\rho$ $(0$--$20\%)$ & $13.94\pm0.0007$ \\
& $1-\rho$ $(80$--$100\%)$ & $15.36\pm0.0007$ \\


\multirow{3}{*}{IV}
& Integrated & $6.27\pm0.0004$ \\
& $1-\rho$ $(0$--$20\%)$ & $4.58\pm0.0005$ \\
& $1-\rho$ $(80$--$100\%)$ & $7.96\pm0.0003$ \\
\hline
\end{tabular}
\caption{Mean charged-particle multiplicities for the multiplicity and
flattenicity classes used in the analysis.\label{tab:eventclass}}
\end{table}

\subsection{Flattenicity-Based Event Classification}
\label{flattenicity}
To investigate the dependence of charge and momentum balance functions on the event topology, collision events are classified using the event-shape observable flattenicity, originally proposed by Ortiz \textit{et al.} \cite{Ortiz:2022mfv}. Unlike conventional multiplicity-based event classifiers, which characterize events solely through the number of produced particles, this multi-differential approach through flattenicity quantifies the spatial uniformity of particle production in the transverse plane. Consequently, it provides additional sensitivity to the relative contributions of soft and hard QCD processes while reducing the bias toward high-multiplicity jet-dominated events.

Following the procedure adopted by the ALICE Collaboration \cite{Ortiz:2022mfv,Ortiz:2022zqr}, flattenicity is evaluated on an event-by-event basis using the charged-particle multiplicity measured within the pseudorapidity acceptance of the V0A ($2.8<\eta<5.1$) and V0C ($-3.7<\eta<-1.7$) detectors over the full azimuthal coverage ($0\leq\phi<2\pi$). The detector acceptance is divided into eight equal intervals in pseudorapidity and eight equal intervals in azimuth, resulting in an $\eta$--$\phi$ grid consisting of 64 elementary cells. The charged-particle multiplicity is determined for each cell, and the event flattenicity is calculated from the event-by-event fluctuations of the multiplicity distribution according to

\begin{equation}
    \rho_{\text{nch}} = 
\frac{
\sqrt{
\sum_i \left(N^{\text{cell},i}_{\text{ch}} - \langle N^{\text{cell}}_{\text{ch}} \rangle \right)^2 / N^2_{\text{cell}}
}
}{
\langle N^{\text{cell}}_{\text{ch}} \rangle
}
\end{equation}

where $N_{\rm ch}^{\rm cell,i}$ denotes the charged-particle multiplicity in the $i^{\rm th}$ elementary cell, $\langle N_{\rm ch}^{\rm cell}\rangle$ is the average charged-particle multiplicity per cell for a given event, and $N_{\rm cell}=64$ is the total number of cells used in the calculation.

The quantity $\rho_{\rm nch}$ represents the relative event-by-event fluctuation of the charged-particle density across the detector acceptance. Following the convention of Ref.~\cite{Ortiz:2022mfv}, the event classifier is defined as $(1-\rho_{\rm nch})$, where values approaching unity correspond to nearly isotropic particle production, while values approaching zero indicate highly non-uniform or jet-like event topologies. Thus, larger values of flattenicity, $(1-\rho_{\rm nch})$ identify events dominated by soft, isotropic particle production, whereas smaller values are associated with events containing localized high-momentum particle clusters originating from hard QCD processes.

To investigate the dependence of the balance functions on event
topology, events are classified into four charged-particle
multiplicity classes and further categorized according to flattenicity within each multiplicity class. The lowest 20\% of the $(1-\rho_{\rm nch})$ distribution, corresponding to the
$0$--$20\%$ percentile, is used to select jet-like events, denoted by $1-\rho$ $(0$--$20\%)$ whereas
the highest 20\%, corresponding to the $80$--$100\%$ percentile,
is used to select isotropic events, represented as $1-\rho$ $(80$--$100\%)$ . The charge balance function,
$B$, and the charge-dependent transverse-momentum balance function, $P_{2}^{\rm CD}$, are evaluated for these event classes to investigate the dependence of charge  and transverse-momentum correlations on event topology. The corresponding average charged-particle multiplicities are summarized in Table~\ref{tab:eventclass}.


\subsection{Charge balance function}
\label{B}
The charge balance function is a two-particle correlation observable that quantifies the correlation between balancing oppositely charged particles produced in high-energy collisions. Owing to local electric charge conservation, charged particles are created in balancing pairs, and the relative separation of these pairs in phase space carries valuable information about the underlying particle production dynamics, including charge creation, time of hadronization, and subsequent transport processes. In the present analysis, the charge balance function is studied as a function of the relative pseudorapidity, $\Delta\eta$, and the relative azimuthal angle, $\Delta\phi$. Following the formalism proposed by Bass, Danielewicz, and Pratt \cite{Bass:2000az, Trainor:2003cn} and further developed in Ref.~\cite{Pratt:2021xvg}, the charge balance function is defined as

\begin{equation}
B = \frac{1}{2} \left[ C_{2}^{(+,-)} + C_{2}^{(-,+)} - C_{2}^{(-,-)} - C_{2}^{(+,+)} \right]
\end{equation}

where $C_{2}^{(+,-)}$, $C_{2}^{(-,+)}$ and $C_{2}^{(+,+)}$, $C_{2}^{(-,-)}$ denote the charge-dependent two-particle correlation functions for unlike-sign and like-sign particle pairs. The unlike-sign combinations contain the correlations between balancing charges, whereas the like-sign combinations primarily account for charge-independent background correlations. Their combination therefore isolates the charge-dependent component associated with local charge conservation.

The two-particle correlation functions are constructed from the normalized same-event and mixed-event pair distributions. The same-event pair distribution is defined as

\begin{equation}
S(\Delta\eta,\Delta\phi)=
\frac{1}{N_{\rm trig}}
\frac{d^{2}N^{\rm pair}}
{d\Delta\eta d\Delta\phi},
\end{equation}

where $N_{\rm trig}$ is the number of trigger particles and $N^{\rm pair}$ denotes the number of trigger--associated particle pairs formed within the same event.

To account for the finite detector acceptance, a mixed-event reference distribution is constructed by pairing particles from different events belonging to the same analysis class. The mixed-event distribution is given by

\begin{equation}
M(\Delta\eta,\Delta\phi)=
\frac{1}{N_{\rm trig}}
\frac{d^{2}N^{\rm mix}}
{d\Delta\eta d\Delta\phi},
\end{equation}

where $N^{\rm mix}$ represents the number of particle pairs obtained from the mixed-event sample. Since particles combined through event mixing are uncorrelated, the mixed-event distribution provides an estimate of the pair acceptance arising from the finite detector acceptance.

The acceptance-corrected two-particle correlation function is obtained as

\begin{equation}
\frac{1}{N_{\text{trig}}} 
\frac{d^{2}N^{\text{pair}}}{d\Delta\eta \, d\Delta\phi} 
= C_{2}(\Delta\eta, \Delta\phi)
= M(0,0) \, \frac{S(\Delta\eta, \Delta\phi)}{M(\Delta\eta, \Delta\phi)}
\end{equation}

where $M(0,0)$ denotes the mixed-event yield at $(\Delta\eta,\Delta\phi)=(0,0)$, corresponding to the maximum pair acceptance. The factor $M(0,0)/M(\Delta\eta,\Delta\phi)$ corrects the measured correlation function for the finite pair acceptance in pseudorapidity and azimuth \cite{Basu:2020ldt,Behera:2025ycg,Behera:2025ymi}.

The charge balance function is subsequently projected onto the $\Delta\eta$ and $\Delta\phi$ directions by integrating over the complementary variable. These one-dimensional projections are used to investigate the dependence of charge balancing on the event topology characterized by flattenicity. Furthermore, the widths of the projected balance functions are extracted using the root-mean-square (RMS) of the distributions, providing a quantitative measure of the relative separation of balancing charge pairs in different event classes.

\subsection{Charge-Dependent Transverse Momentum Balance Function}
\label{P2}
While the charge balance function probes the spatial correlation of balancing electric charges, the charge-dependent transverse momentum balance function, $P_{2}^{\rm CD}$, provides complementary information on the correlation of the transverse momenta of charged particle pairs. This observable is sensitive to the covariance of particle transverse momenta relative to the inclusive mean transverse momentum and therefore provides insight into the mechanisms of momentum transport, hadronization, and particle production in high-energy collisions \cite{STAR:2005vxr,Basu:2020ldt}. In the present analysis, $P_{2}^{\rm CD}$ is investigated as a function of the relative pseudorapidity ($\Delta\eta$) and the relative azimuthal angle ($\Delta\phi$).

The dimensionless transverse momentum correlator is defined as

\begin{equation}
P_{2}(\Delta\eta,\Delta\phi)=
\frac{
\left\langle
\Delta p_{{\rm T},1}
\Delta p_{{\rm T},2}
\right\rangle
(\Delta\eta,\Delta\phi)
}
{
\langle p_{\rm T}\rangle^{2}
},
\end{equation}

where $\Delta p = p_{T}-\langle p_{T} \rangle$ represents the deviation of the transverse momentum of a particle from the inclusive mean transverse momentum, and the subscripts 1 and 2 denote the trigger and associated particles, respectively.

The differential transverse momentum correlator is calculated as

\begin{equation}
\langle \Delta p_{\rm T} \Delta p_{\rm T} \rangle (\Delta \eta, \Delta \phi)
= \frac{ \int_{p_{\rm T,\min}}^{p_{\rm T,\max}} \Delta p_{\rm T,1} \, \Delta p_{\rm T,2} \,
C_{2}'(\mathbf{p_{1}}, \mathbf{p_{2}}) \, dp_{T,1} \, dp_{T,2} }
{ \int_{p_{\rm T,\min}}^{p_{\rm T,\max}} C_{2}'(\mathbf{p_{1}}, \mathbf{p_{2}}) \,
dp_{T,1} \, dp_{T,2} } ,
\end{equation}

where $C_{2}'(\mathbf{p_{1}}, \mathbf{p_{2}})$ denotes the differential two-particle density as a function of the transverse momenta of the trigger and associated particles.

The transverse momentum correlator is evaluated independently for the four charge combinations $(+,-)$, $(-,+)$, $(+,+)$, and $(-,-)$. The charge-dependent transverse momentum balance function is then constructed according to
\begin{equation}
P_{2}^{\text{CD}} = \frac{1}{2} \left[ P_{2}^{(+,-)} + P_{2}^{(-,+)} - P_{2}^{(-,-)} - P_{2}^{(+,+)} \right]
\end{equation}
The charge-dependent combination suppresses charge-independent contributions and isolates the momentum correlations associated with balancing charged particles. Positive values of $P_{2}^{\rm CD}$ correspond to a positive covariance between the transverse momenta of balancing particle pairs, whereas negative values indicate an anti-correlation with respect to the inclusive mean transverse momentum.

The charge-dependent transverse momentum balance function is projected onto the $\Delta\eta$ and $\Delta\phi$ directions and analyzed for different flattenicity classes. The widths of the projected distributions are subsequently determined using the root-mean-square (RMS), allowing a quantitative comparison of momentum balancing across different event topologies.

\subsection{Projection Procedure and Width Extraction}

The two-dimensional charge balance function and charge-dependent transverse momentum balance function are evaluated over the ranges $-1.6 \leq \Delta\eta \leq 1.6$ and $-\pi/2 \leq \Delta\phi \leq 3\pi/2$. To facilitate a quantitative comparison of the correlation structures, one-dimensional projections are constructed along the $\Delta\eta$ and $\Delta\phi$ directions. The $B(\Delta\eta)$ and $P_{2}^{\rm CD}(\Delta\eta)$ distributions are obtained by projecting the corresponding two-dimensional correlation functions over the interval $-\pi/2 \leq \Delta\phi \leq \pi/2$, while $B(\Delta\phi)$ and $P_{2}^{\rm CD}(\Delta\phi)$ are projected over $-1.0 \leq \Delta\eta \leq 1.0$. The projections are normalized by the number of bins included in the respective projection intervals.

To quantify the extent of the correlations, the widths of the projected distributions are determined using the root-mean-square (RMS) evaluated over the ranges $-1.0 \leq \Delta\eta \leq 1.0$ and $-\pi/2 \leq \Delta\phi \leq \pi/2$. The RMS width is calculated as

\begin{equation}
\sigma =
\left[
\frac{
\displaystyle\sum_i
O(\Omega_i),
\Omega_i^{,2}
}{
\displaystyle\sum_i
O(\Omega_i)
}
\right]^{1/2},
\label{RMSwidths}
\end{equation}

where $O(\Omega_i)$ denotes the value of the projected observable, either the charge balance function or the charge-dependent transverse momentum balance function, evaluated at the bin center $\Omega_i$, with $\Omega$ representing either $\Delta\eta$ or $\Delta\phi$ \cite{Behera:2025ycg,Behera:2025ymi,Roy:2026ydm}.

The RMS width provides a quantitative measure of the spatial extent of the correlation function. A narrower width indicates that balancing charge or momentum pairs remain more closely correlated in phase space, whereas a broader width corresponds to a larger separation between correlated particles. The extracted widths are subsequently compared for different flattenicity classes to investigate the influence of event topology on charge and momentum balancing in proton--proton collisions.

\section{Results and Discussion}
\label{results}
\subsection{Flattenicity Distribution}
\begin{figure}[htbp]
    \centering
    \includegraphics[width=1.0\linewidth]{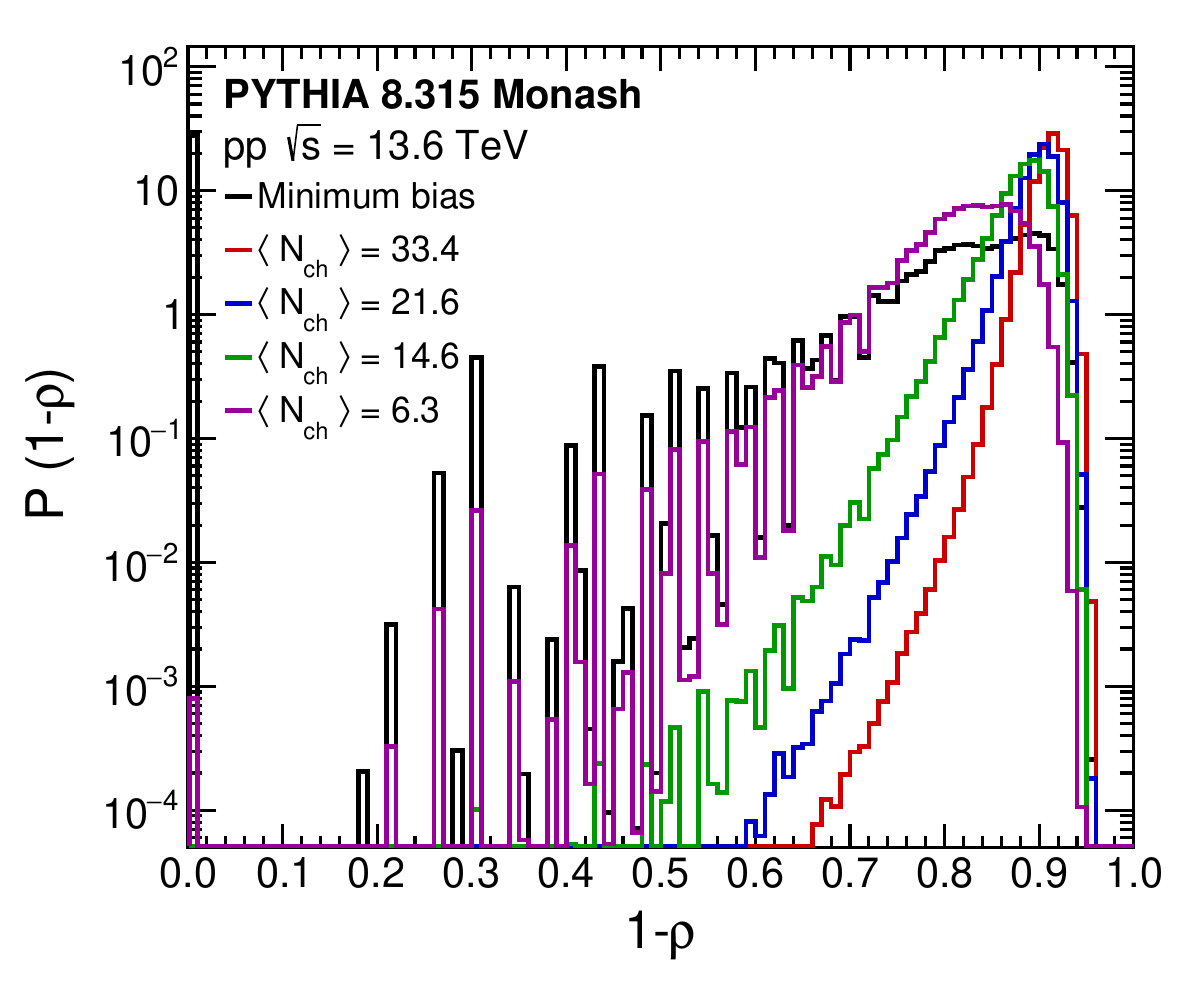}
    \caption{Normalized event-by-event flattenicity ($1-\rho$) distributions obtained from PYTHIA~8.315 simulations of $pp$ collisions at $\sqrt{s}=13.6$~TeV for minimum-bias events and different charged-particle multiplicity classes.}
    \label{fig:FlatDist}
\end{figure}

Figure~\ref{fig:FlatDist} shows the normalized flattenicity distributions, $(1-\rho)$, for minimum-bias $pp$ collisions at $\sqrt{s}=13.6$~TeV and for different charged-particle multiplicity classes simulated using \texttt{PYTHIA~8}.


The flattenicity distributions exhibit a clear dependence on charged-particle multiplicity, with higher-multiplicity classes showing a shift towards larger values of $(1-\rho)$. The minimum-bias distribution reflects the inclusive event sample and lies between the distributions of different multiplicity classes.
This suggests that high-multiplicity proton-proton collisions are characterized by a more uniform distribution of charged particles in the transverse plane (isotropic events), which corresponds to higher $(1-\rho)$ values. Low-multiplicity events have larger relative fluctuations in particle density across the V0A and V0C acceptance regions, leading to smaller $(1-\rho)$ values and broader flattenicity distributions. These larger fluctuations are caused mostly by the small number of produced particles and the increased likelihood of localized particle production (jetty events) in low-multiplicity events. This demonstrates that flattenicity provides a useful event-shape classification beyond conventional multiplicity-based event selection. While charged-particle multiplicity primarily characterizes the overall event activity, flattenicity additionally accounts for the spatial distribution of particle production across the forward $\eta$--$\phi$ acceptance. This allows events with similar multiplicities but different particle-density patterns to be distinguished, providing sensitivity to variations in event topology and the underlying particle-production activity. In particular, the use of the forward V0A and V0C regions makes flattenicity complementary to midrapidity observables and provides a useful handle for investigating the dependence of particle correlations on the global event structure. The sensitivity of flattenicity to multiparton interactions and its reduced susceptibility to certain multiplicity-selection biases further motivate its use in pp collisions.



The flattenicity distributions for each charged-particle multiplicity class are used to select the lowest 20\% and highest 20\% of events, corresponding to jet-like and isotropic event topologies, respectively. The charge balance function, $B$, and the charge-dependent transverse-momentum balance function, $P_{2}^{\rm CD}$, are then studied for these event selections to investigate their dependence on event topology at different charged-particle multiplicities.

\subsection{Two-Dimensional Distributions of $B(\Delta\eta,\Delta\phi)$}
\label{2D_B}

\begin{figure*}[htb]
    \centering

    \begin{subfigure}{0.45\textwidth}
        \centering
        \includegraphics[width=\linewidth]{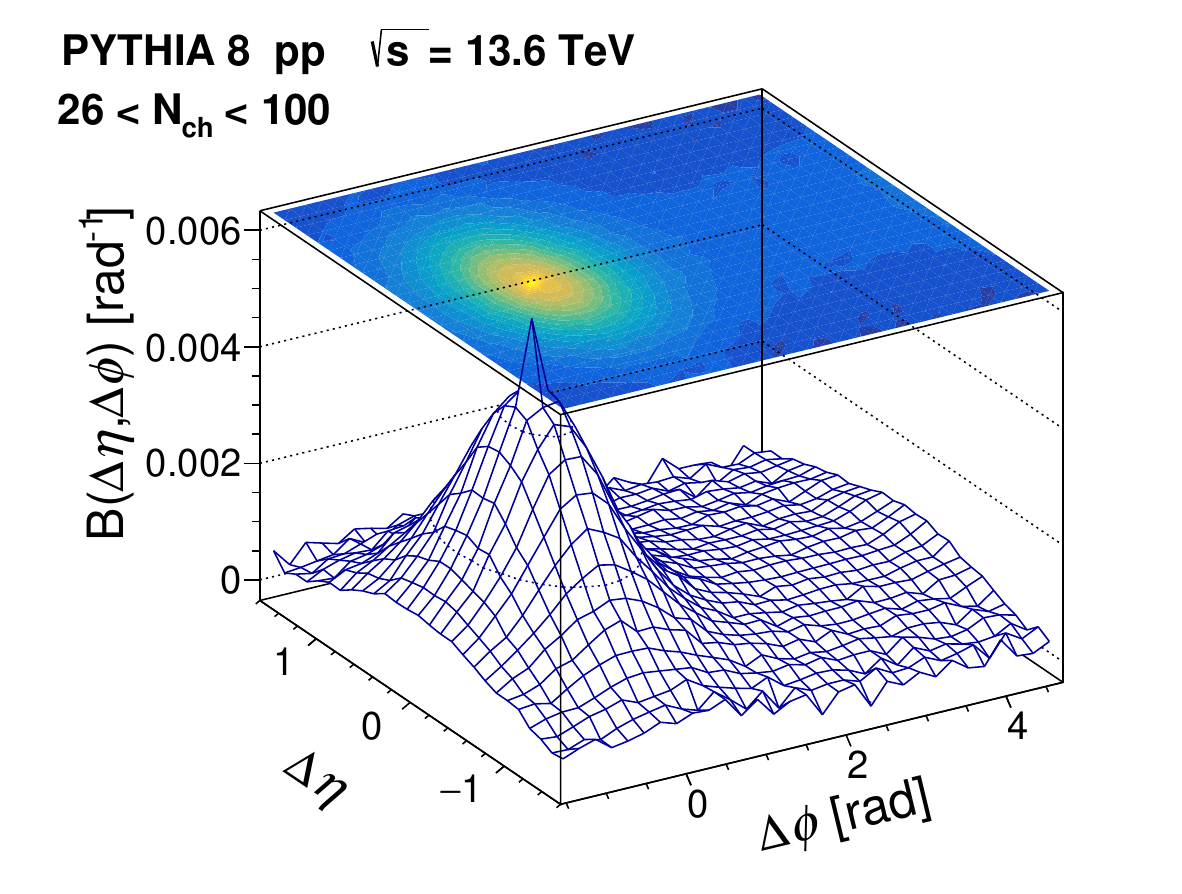}
        \caption{Jetty $1-\rho$ $(0$--$20\%)$}
        \label{fig:a}
    \end{subfigure}
    \hfill
    \begin{subfigure}{0.45\textwidth}
        \centering
        \includegraphics[width=\linewidth]{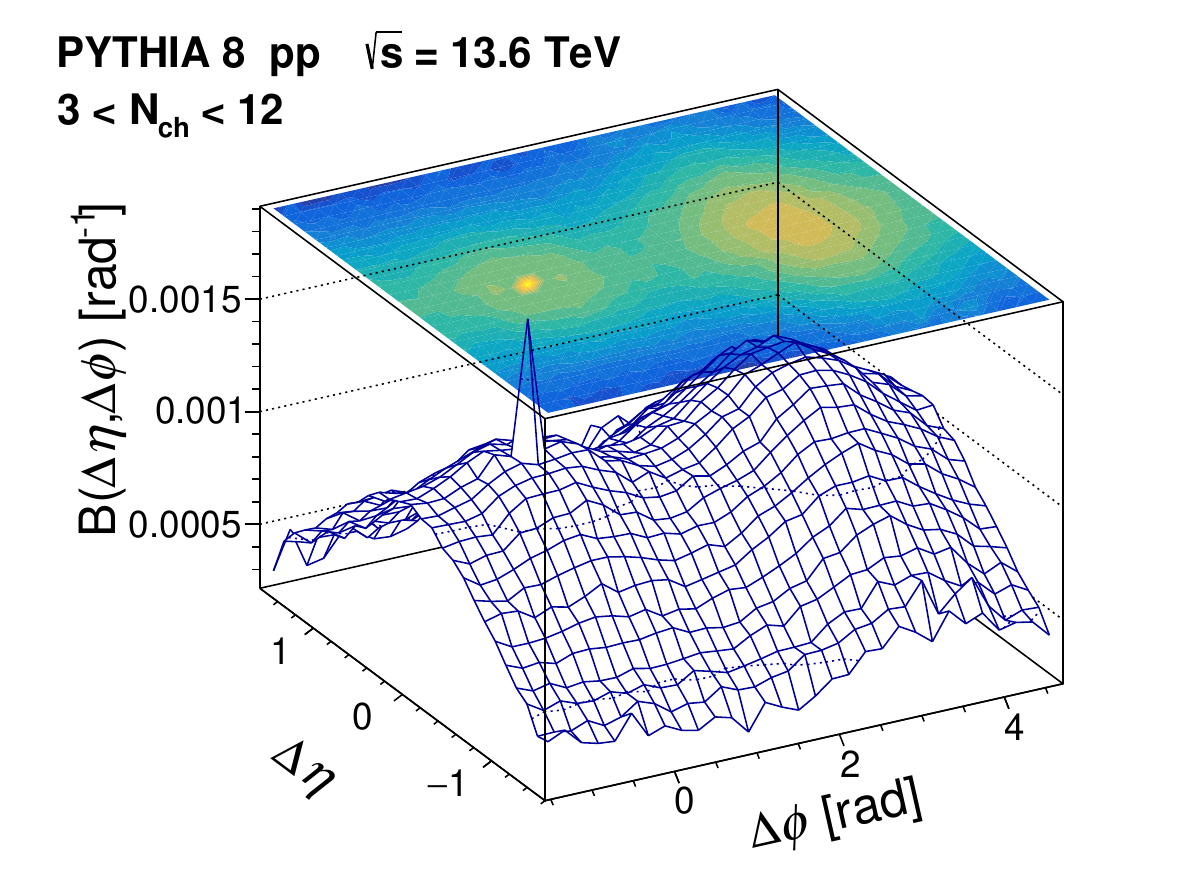}
        \caption{Jetty $1-\rho$ $(0$--$20\%)$}
        \label{fig:b}
    \end{subfigure}

    \vspace{0.5cm}

    \begin{subfigure}{0.45\textwidth}
        \centering
        \includegraphics[width=\linewidth]{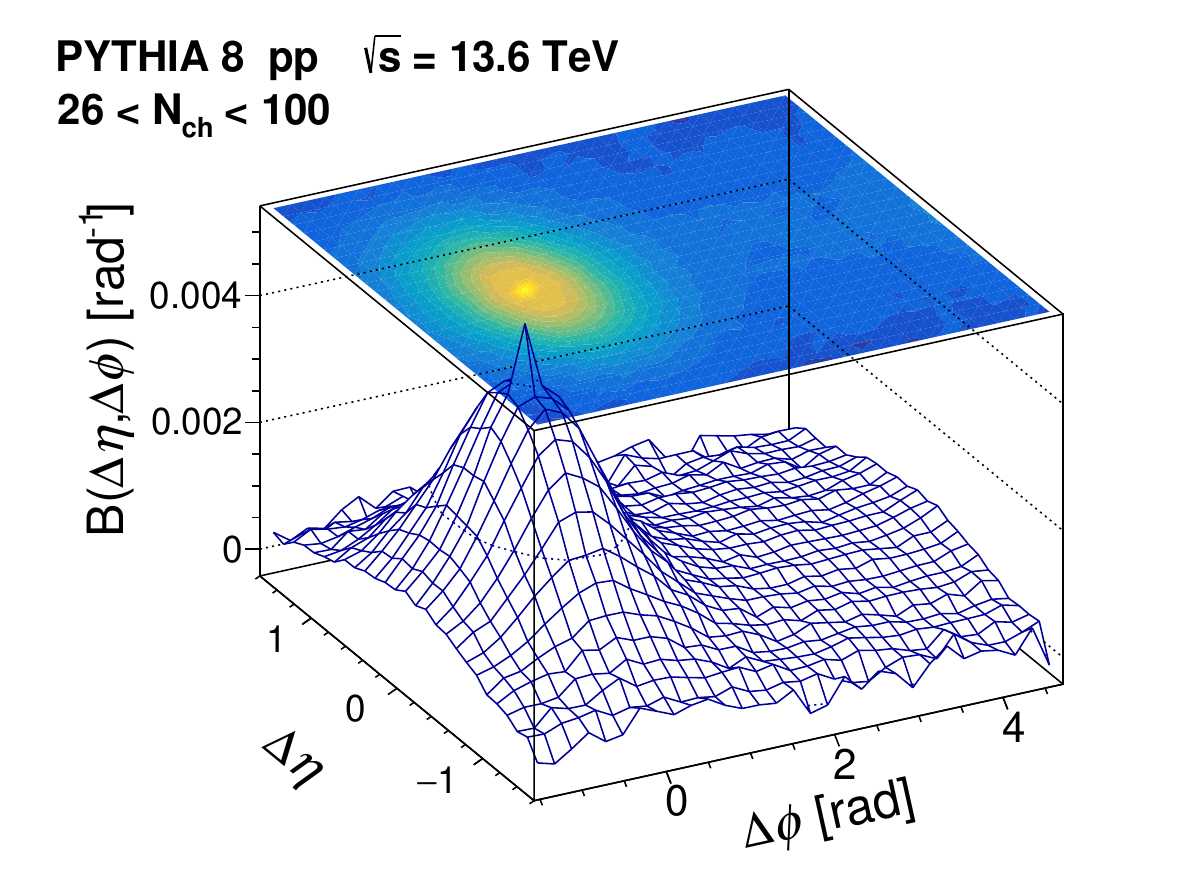}
        \caption{Isotropic $1-\rho$ $(80$--$100\%)$}
        \label{fig:c}
    \end{subfigure}
    \hfill
    \begin{subfigure}{0.45\textwidth}
        \centering
        \includegraphics[width=\linewidth]{B_Flat0_Mult4.pdf}
        \caption{Isotropic $1-\rho$ $(80$--$100\%)$}
        \label{fig:d}
    \end{subfigure}
  \caption{Two-dimensional charge balance function, $B(\Delta\eta,\Delta\phi)$, for (a) the highest- and (b) lowest-multiplicity events in the $1-\rho$ $(0$--$20\%)$ class, corresponding to jet-like events, and (c) the highest- and (d) lowest-multiplicity events in the $1-\rho$ $(80$--$100\%)$ class, corresponding to isotropic events.
}
    \label{fig:B2D_Flat}
\end{figure*}

Figure \ref{fig:B2D_Flat} shows the two-dimensional charge balance function, $B(\Delta\eta,\Delta\phi)$, for the jet-like, $1-\rho$ $(0$--$20\%)$ and isotropic, $1-\rho$ $(80$--$100\%)$ flattenicity classes, for the lowest and the highest multiplicity events. For both multiplicity classes, the balance function exhibits a pronounced near-side peak around \((\Delta\eta,\Delta\phi)\approx(0,0)\), indicating strong short-range charge correlations between balancing pairs. This suggests that oppositely charged particles are preferentially produced with small separations in pseudorapidity and azimuth.




In addition to the prominent near-side peak, a broad away-side correlation structure is observed around \(\Delta\phi\approx\pi\), particularly in the lowest multiplicity class, extending over a broad range of \(\Delta\eta\). This away-side structure is primarily associated with the approximately back-to-back emission of particles from the underlying partonic scattering and fragmentation, with transverse-momentum conservation contributing to the characteristic azimuthal correlation. Its stronger presence at low multiplicity can be attributed to the relatively larger contribution of jet-like particle production in these events. With increasing multiplicity, the contribution from softer particle-production mechanisms and multiparton interactions becomes more significant, increasing the non-jet component and thereby diluting and broadening the away-side jet correlation. A similar multiplicity-dependent weakening of the away-side balance-function structure has been observed in \texttt{PYTHIA8} studies \cite{Behera:2025ymi} and in experimental measurements \cite{ALICE:2015nuz} of pp collisions.

These variations are more apparent in the projected balance functions and corresponding RMS widths mentioned in sections \ref{1D_B} and \ref{B_widths}.

\subsection{One-Dimensional Projections of $B(\Delta\eta,\Delta\phi)$}
\label{1D_B}

\begin{figure*}[htb]
	\centering 
	\includegraphics[width=1.0\textwidth, angle=0]{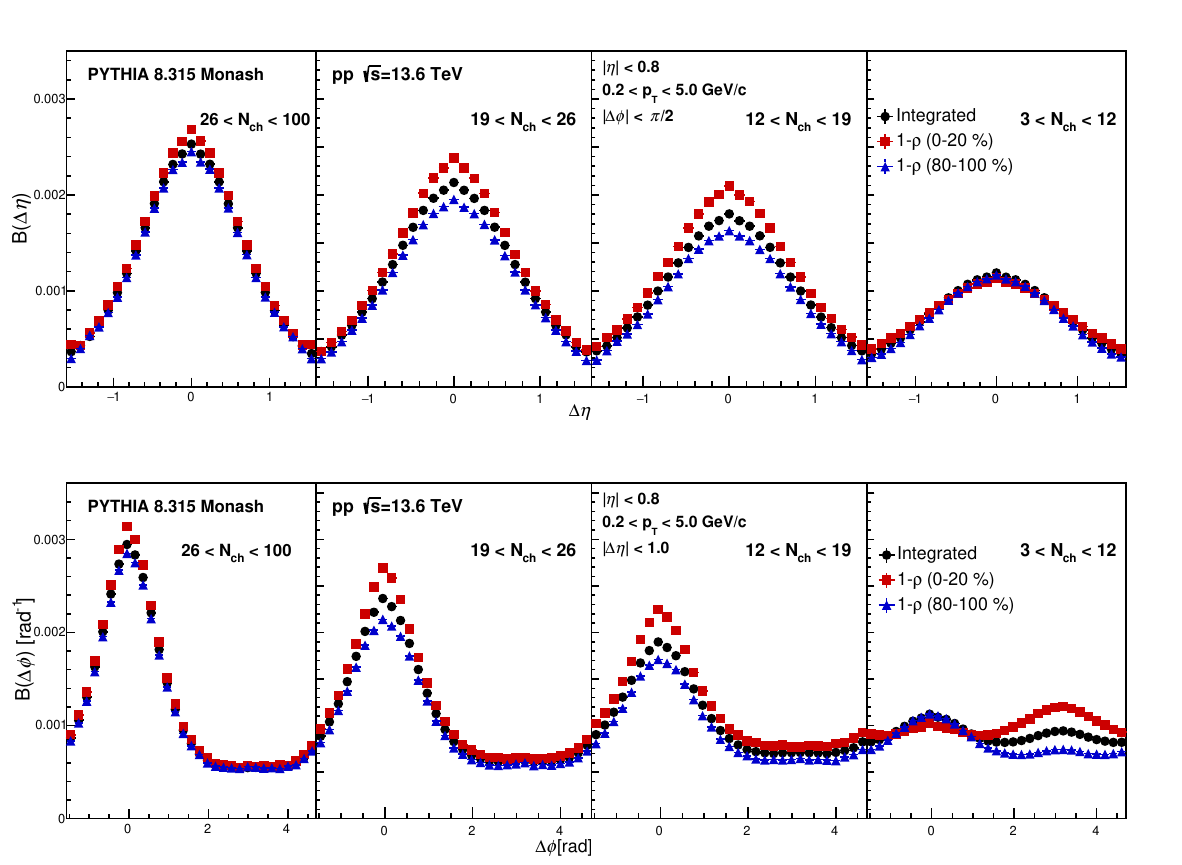}	
	\caption{One-dimensional pseudorapidity (top panel) and azimuthal (bottom panel) projections of the charge balance function, $B(\Delta\eta)$, for different charged-particle multiplicity intervals and flattenicity classes.} 
	\label{fig:B_eta}
\end{figure*}

Fig.~\ref{fig:B_eta} depict one-dimensional projections of the charge balance function in the pseudorapidity and azimuthal directions for various multiplicity intervals and two extreme flattenicity classes. For comparison we have plotted the balance function for the integrated class. As already explained in section~\ref{2D_B}, the distributions show significant near-side peaks at $\Delta \eta = 0$ and $\Delta \phi = 0$, showing that balancing charge pairs are primarily created close together in phase space. The balance functions further exhibit a clear multiplicity dependence in both pseudorapidity and azimuthal projections. With increasing charged-particle multiplicity, the near-side peak becomes progressively stronger and at the same time  narrower, indicating an increasing localization of balancing-charge correlations around \(\Delta\eta\approx0\) and \(\Delta\phi\approx0\). This behavior is consistent with the multiplicity-dependent narrowing of balance functions reported in \texttt{PYTHIA8} and experimental  studies \cite{CMS:2023sua,ALICE:2015nuz,Manea:2024qgd}.In the \texttt{PYTHIA~8} framework, the observed narrowing of the
correlation structure may be related to the interplay of multiparton interactions (MPI) and hadronization dynamics, including colour reconnection (CR), which might rearrange colour connections between partons and can influence the final-state particle correlations \cite{Behera:2025ymi,Roy:2026ydm}.

In the azimuthal projection (Fig.~\ref{fig:B_eta} bottom panel), a broad away-side structure around ($\Delta\phi\approx\pi$) is more prominent at lower multiplicities, while it becomes weaker and broader with increasing multiplicity. A similar evolution has been observed in \texttt{PYTHIA8} calculations of particle-identified balance functions \cite{Behera:2025ycg}, where the low-multiplicity events exhibit a pronounced away-side component associated with approximately back-to-back particle (dijet) production \cite{ALICE:2023lyr}, whereas the near-side contribution becomes increasingly prominent at higher multiplicities. This behavior may indicate a change in the relative contributions of jet-related and softer particle-production mechanisms with increasing multiplicity. For all multiplicity classes, jet-like ($1-\rho$ $(0$--$20\%)$) flattenicity events show the largest correlation strength reflecting more localized charge balancing in jet-like events \cite{Behera:2026bso} unaffected by longitudinal diffusion. However, isotropic ($1-\rho$ $(80$--$100\%)$) events, dominated by soft processes, produce comparatively smaller peak amplitudes. The integrated distributions reside between the two extreme flattenicity classes. The dependence of the charge balance function on flattenicity varies across the charged-particle multiplicity classes in both
$\Delta\eta$ and $\Delta\varphi$ projections. The differences
between jet-like and isotropic events are more pronounced in the
intermediate-multiplicity classes, whereas comparatively smaller
differences are observed in the highest- and lowest-multiplicity
classes. This behaviour suggests that the sensitivity of
charge-dependent correlations to event topology depends on
charged-particle multiplicity.

\subsection{Multiplicity and Flattenicity Dependence of the $B(\Delta\eta,\Delta\phi)$ Widths}
\label{B_widths}

The RMS widths $\sigma(\Delta\eta)$ and $\sigma(\Delta\phi)$ of the projected charge balance functions, calculated using Eq.~\ref{RMSwidths}, are presented in Fig.~\ref{fig:Widths} as a function of the average charged-particle multiplicity for the integrated event sample and the two extreme flattenicity classes. Both $\sigma(\Delta\eta)$ and $\sigma(\Delta\phi)$ exhibit a systematic decrease with increasing charged-particle multiplicity, indicating that the balancing-charge correlations become increasingly localized in relative pseudorapidity and azimuth at higher multiplicities as explained in section~\ref{1D_B}. Such decreasing pattern for the integrated class has already been observed in \cite{ALICE:2015nuz,ALICE:2013tla} are also consistent with the two extreme flattenicity classes. 

Both $\sigma(\Delta\eta)$ and $\sigma(\Delta\phi)$  show a moderate dependence on flattenicity in the lowest multiplicity class. Although small differences are observed among the three event-shape classes, these differences decrease with increasing charged-particle multiplicity, and the widths become nearly identical in the highest multiplicity interval. The relatively small separation between the flattenicity classes suggests that the multiplicity dependence of the balance-function width is more pronounced than its event-shape dependence. Within the \texttt{PYTHIA8} framework, the observed narrowing may be influenced by changes in the relative contributions of multiparton interactions and color reconnection \cite{ALICE:2018jco}, which modify the color topology and hadronization pattern and can consequently affect the correlations of balancing charges.


\subsection{Two-Dimensional Distributions of $P_{2}^{\rm CD}(\Delta\eta,\Delta\phi)$} 
\label{2D_P2}

\begin{figure*}[htb]
    \centering

    \begin{subfigure}{0.45\textwidth}
        \centering
        \includegraphics[width=\linewidth]{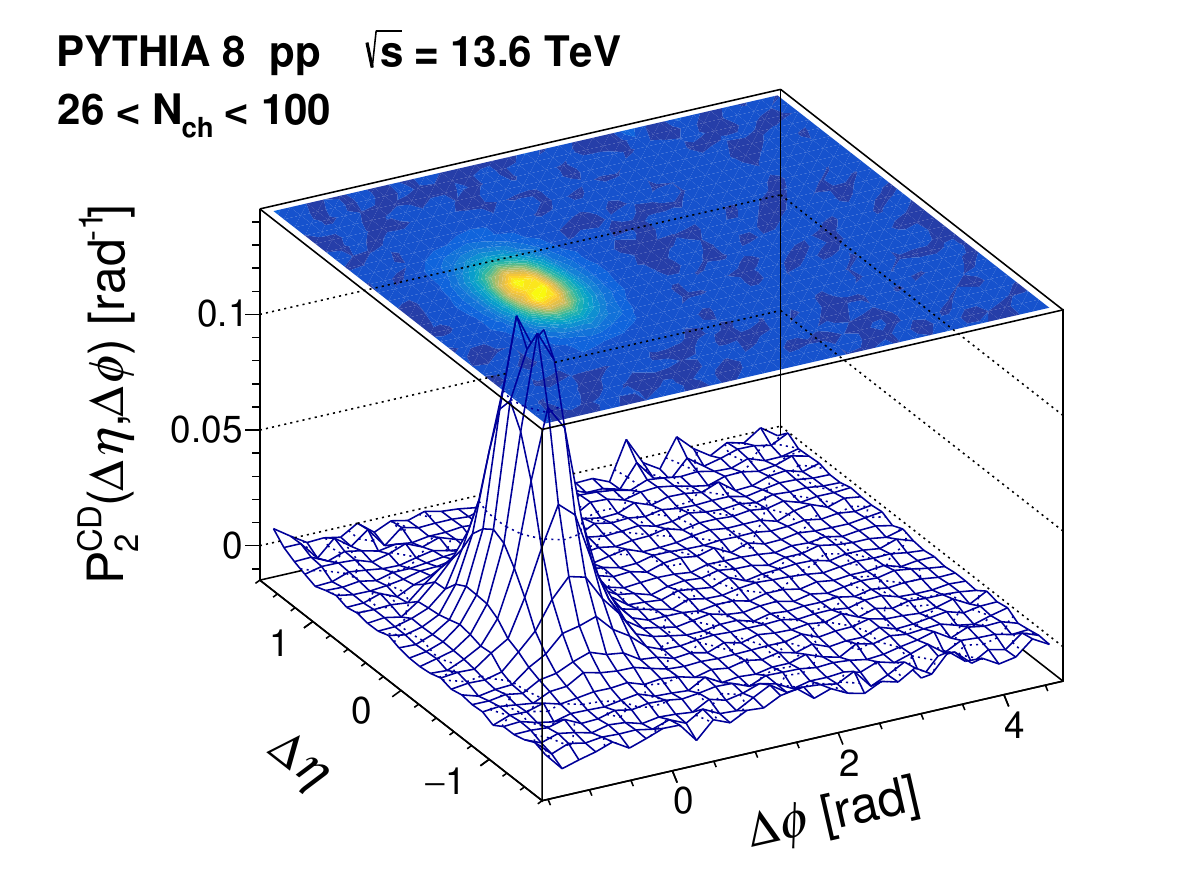}
        \caption{Jetty $1-\rho$ $(0$--$20\%)$}
        \label{fig:a}
    \end{subfigure}
    \hfill
    \begin{subfigure}{0.45\textwidth}
        \centering
        \includegraphics[width=\linewidth]{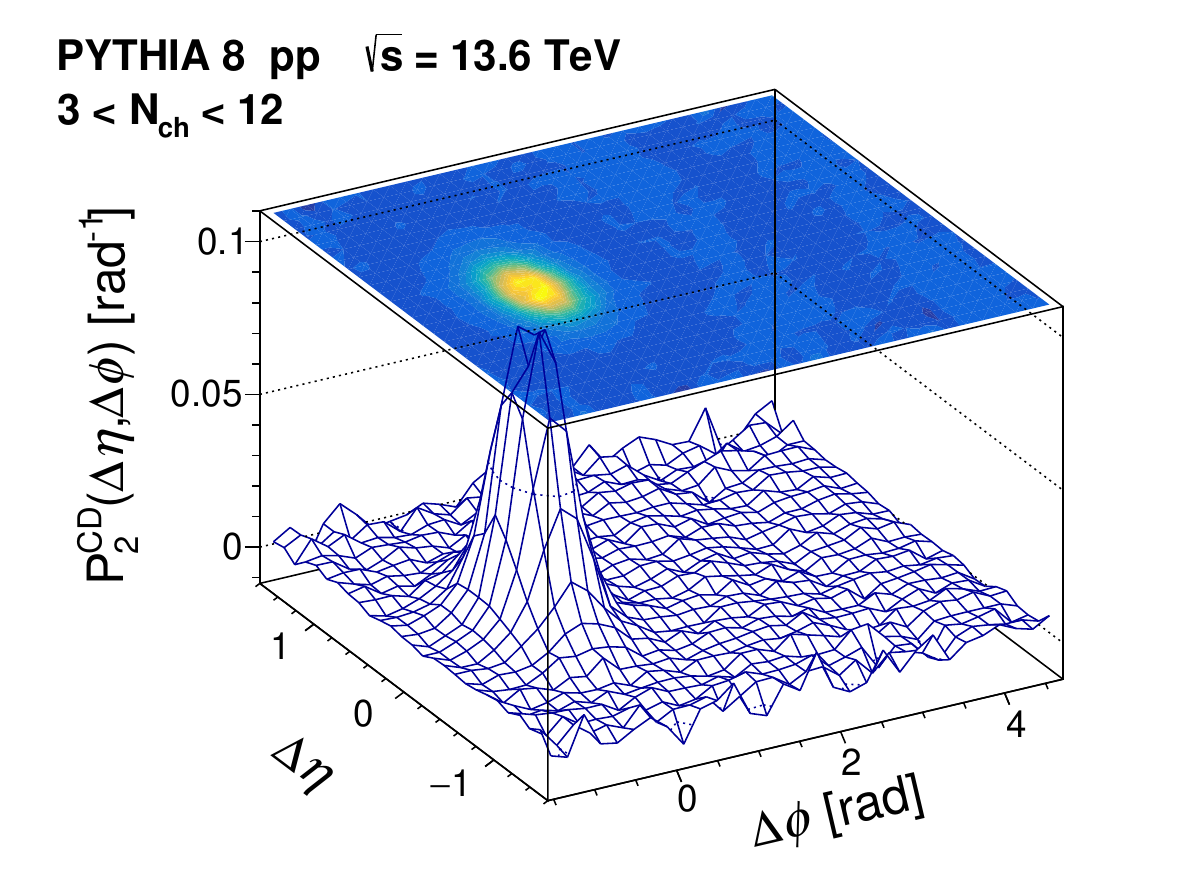}
        \caption{Jetty $1-\rho$ $(0$--$20\%)$}
        \label{fig:b}
    \end{subfigure}

    \vspace{0.5cm}

    \begin{subfigure}{0.45\textwidth}
        \centering
        \includegraphics[width=\linewidth]{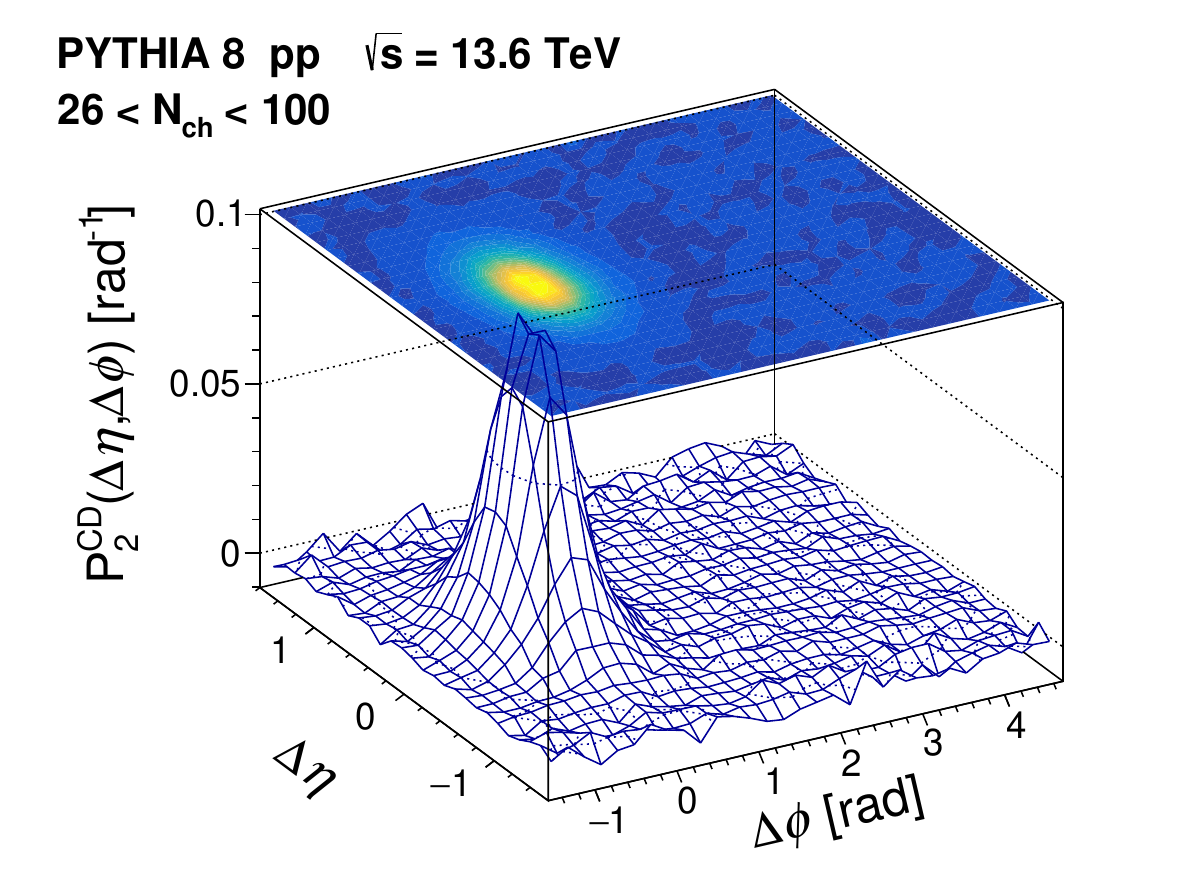}
        \caption{Isotropic $1-\rho$ $(80$--$100\%)$}
        \label{fig:c}
    \end{subfigure}
    \hfill
    \begin{subfigure}{0.45\textwidth}
        \centering
        \includegraphics[width=\linewidth]{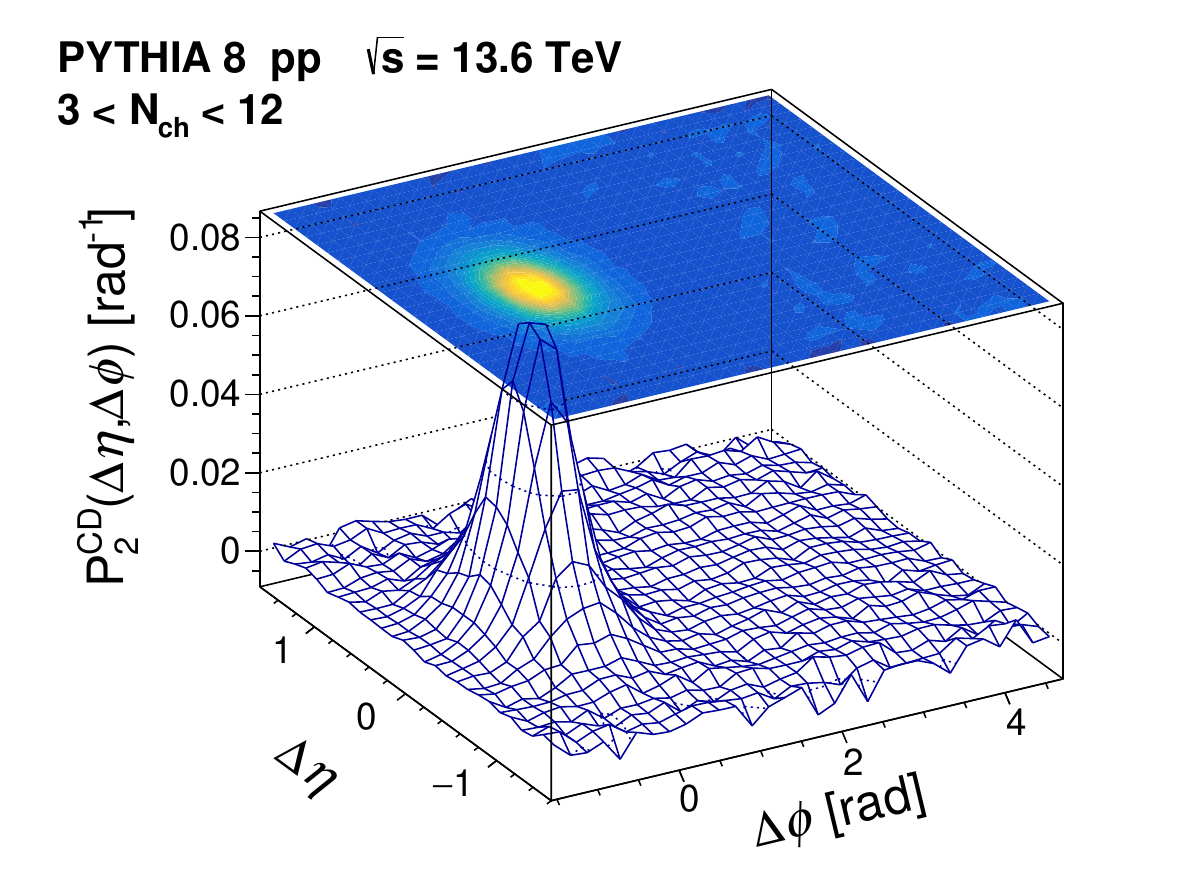}
        \caption{Isotropic $1-\rho$ $(80$--$100\%)$}
        \label{fig:d}
    \end{subfigure}

   \caption{Two-dimensional charge-dependent transverse momentum balance function, $P_{2}^{\rm CD}(\Delta\eta,\Delta\phi)$, for (a) the highest- and (b) lowest-multiplicity events in the $1-\rho$ $(0$--$20\%)$ class, corresponding to jet-like events, and (c) the highest- and (d) lowest-multiplicity events in the $1-\rho$ $(80$--$100\%)$ class, corresponding to isotropic events.
}
    \label{fig:P2Flat}
\end{figure*}

Fig.~\ref{fig:P2Flat} presents the two-dimensional charge-dependent transverse momentum balance function, $P_{2}^{\rm CD}(\Delta\eta,\Delta\phi)$, for $1-\rho$ $(0$--$20\%)$ (jet-like) and $1-\rho$ $(80$--$100\%)$ (isotropic) flattenicity classes, respectively, for the lowest and the highest multiplicity events. For both the multiplicity intervals and flattenicity classes, the correlation function exhibits a pronounced near-side peak centered around $(\Delta\eta,\Delta\phi)\approx(0,0)$, indicating that balancing particle pairs with correlated transverse momenta are primarily produced close to each other in phase space. Away from the near-side region, the correlation strength decreases rapidly and remains close to zero, indicating that the charge-dependent transverse-momentum correlation is predominantly localized around \((\Delta\eta,\Delta\phi)\approx(0,0)\). This localized structure also reflects short-range charge-dependent correlations and is consistent with previous \texttt{PYTHIA}-based studies \cite{Roy:2026ydm}.

\subsection{One-Dimensional Projections of $P_{2}^{\rm CD}(\Delta\eta,\Delta\phi)$}

\begin{figure*}[htb]
	\centering 
	\includegraphics[width=1.0\textwidth, angle=0]{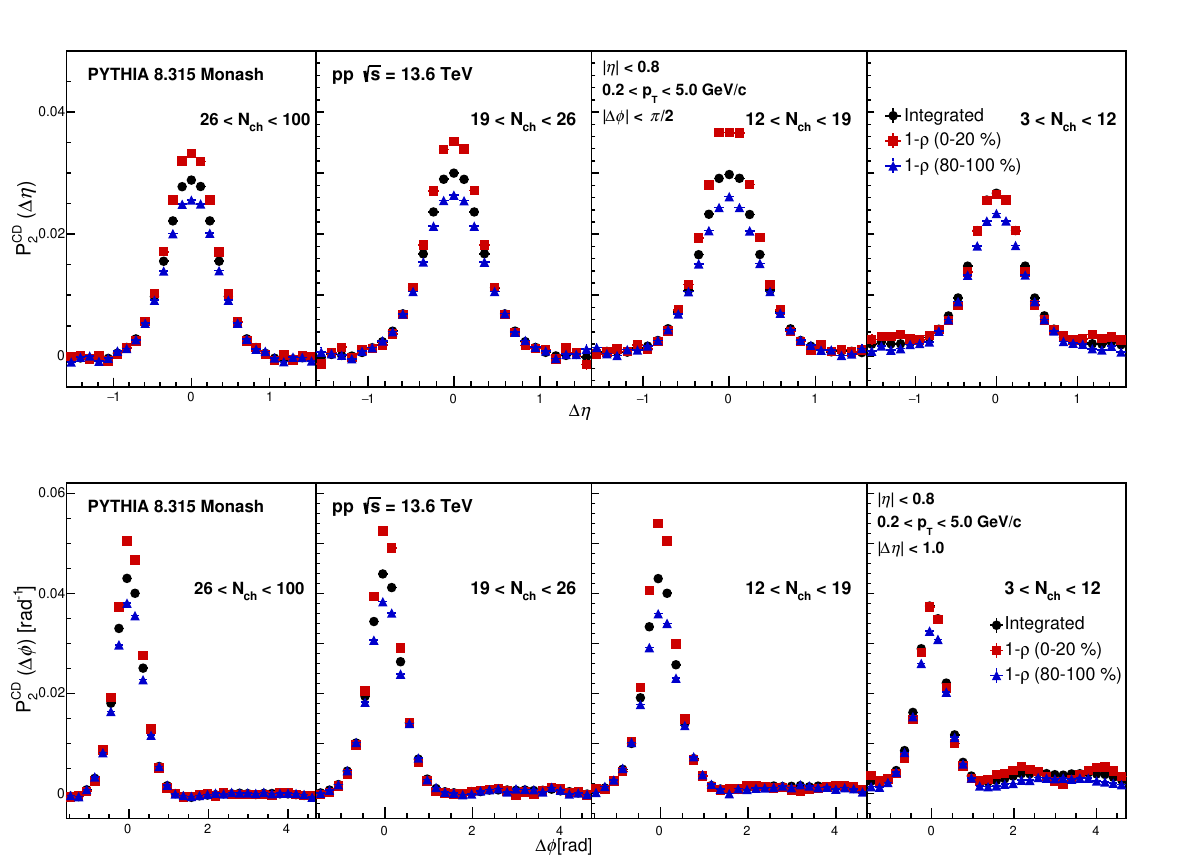}	
	\caption{One-dimensional pseudorapidity (top panel) and azimuthal (bottom panel) projections of the charge-dependent transverse momentum balance function, $P_{2}^{\rm CD}(\Delta\eta)$, for different charged-particle multiplicity intervals and flattenicity classes.} 
	\label{fig:P2ProjectionEta}%
\end{figure*}

The one-dimensional projections of the charge-dependent transverse momentum balance function along the pseudorapidity (top panel) and azimuthal (bottom panel) directions are presented in Fig.~\ref{fig:P2ProjectionEta}, respectively, for different multiplicity intervals and the two extreme flattenicity classes. Both \(P_2^{CD}(\Delta\eta)\) and \(P_2^{CD}(\Delta\phi)\) exhibit a prominent peak around \(\Delta\eta=0\) and \(\Delta\phi=0\), respectively, indicating that charge-dependent transverse-momentum correlations are predominantly localized within a limited region of phase space as already explained in section~\ref{2D_P2}. As shown in Fig.~\ref{fig:P2ProjectionEta} the correlation strength decreases rapidly with increasing separation from the near-side region, while the away-side remains largely structureless. However, a weak away-side structure gradually emerges towards the lower multiplicity intervals, possibly reflecting the increasing relative contribution of back-to-back particle production associated with jet fragmentation in lower-multiplicity events. A similar behavior has been observed in \texttt{PYTHIA8}-related studies \cite{Behera:2025ymi}, where jetty and isotropic events were classified using spherocity. Further, both \(P_2^{CD}(\Delta\eta)\) and \(P_2^{CD}(\Delta\phi)\) exhibit a significant dependence on flattenicity.
This dependence becomes more prominent towards higher multiplicity classes, where a clear separation between the different event-shape selections is observed. For all multiplicity intervals, the jet-like \(1-\rho\) (0–20\%) class exhibits the largest near-side peak, whereas the isotropic \(1-\rho\) (80–100\%) class shows a comparatively smaller peak, with the integrated sample lying between the two.

\subsection{Multiplicity and Flattenicity Dependence of the $P_{2}^{\rm CD}(\Delta\eta,\Delta\phi)$ Widths}

\begin{figure}[h]
    \centering

        \centering
        \includegraphics[width=\linewidth]{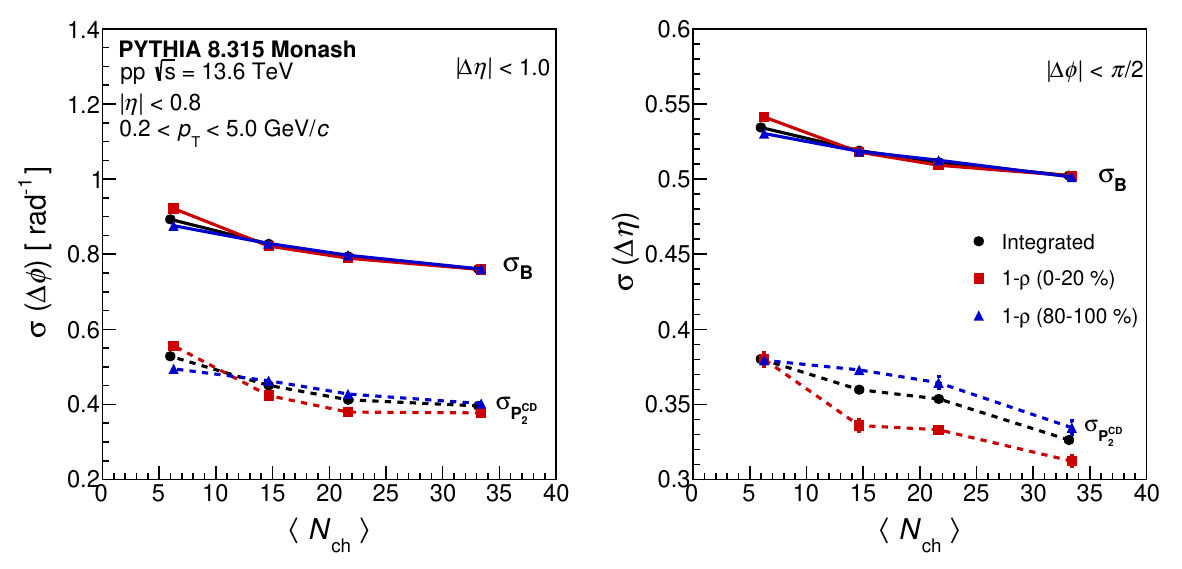}
      \caption{Root-mean-square (RMS) widths of the projected charge balance function $B$ (solid lines) and charge-dependent transverse-momentum balance
      function $P_{2}^{\rm CD}$ (dashed lines) in azimuth, $\sigma(\Delta\phi)$ (left panel), and pseudorapidity, $\sigma(\Delta\eta)$ (right panel), as a function of the average charged-particle multiplicity for the integrated event sample and the two extreme flattenicity classes.}
    \label{fig:Widths}
\end{figure}

Fig.~\ref{fig:Widths} presents the RMS widths $\sigma_{P_{2}^{CD}}(\Delta\eta)$ and $\sigma_{P_{2}^{CD}}(\Delta\phi)$  of the projected charge-dependent transverse momentum balance functions calculated using Eq. ~\ref{RMSwidths}, along the pseudorapidity and azimuthal directions for the integrated event sample and the two extreme flattenicity classes. Both $\sigma_{P_{2}^{CD}}(\Delta\eta)$ and $\sigma_{P_{2}^{CD}}(\Delta\phi)$ decrease systematically with increasing charged-particle multiplicity indicating that, similar to charge balance function, transverse momentum balance function also becomes increasingly localized in high-multiplicity proton--proton collisions. It is also seen from Fig.~\ref{fig:Widths}, the widths $\sigma_{P_{2}^{CD}}(\Delta\eta)$ and $\sigma_{P_{2}^{CD}}(\Delta\phi)$ exhibit a pronounced flattenicty dependence. The jet-like $1-\rho$ $(0$--$20\%)$ flattenicity class exhibits the smallest RMS widths, whereas the isotropic $1-\rho$ $(80$--$100\%)$ events produce the largest widths, with the integrated event sample lying between the two extreme event classes. The narrower widths observed for jet-like events are consistent with a stronger contribution from jet-fragmentation processes. In contrast, the broader widths observed for isotropic events may reflect a relatively larger contribution from soft particle production and underlying event activity, including multiparton interactions (MPI) and colour reconnection (CR). These mechanisms believed to modify the relative composition and correlations of the produced particles and may therefore influence the transverse-momentum balance-function widths \cite{Behera:2025ymi}. The flattenicity dependence is more pronounced for $P_{2}^{\rm CD}$ than for $B$, demonstrating the stronger sensitivity of $P_{2}^{\rm CD}$ to the underlying event topology.

\subsection{Comparison of the $B$ and $P_2^{CD}$} \label{subsec:BF_P2CD_comparison}

We can see from the 2D plots in Fig.~\ref{fig:B2D_Flat} and Fig.~\ref{fig:P2Flat} that the correlation structures of $B$ and $P_2^{CD}$ are visibly different which can be attributed to their sensitivity to particle-pair correlations in different ways. To quantify these differences, we obtained one-dimensional projections of both observables along the pseudorapidity (\(\Delta\eta\)) and azimuthal (\(\Delta\phi\)) directions and calculated their corresponding RMS widths which have been plotted in Fig.~\ref{fig:Widths} as a function of average charged particle multiplicity for different flattenicity classes. As shown in Fig.~\ref{fig:Widths}, the RMS widths of $P_2^{CD}$ are systematically smaller than the widths of $B$ in both pseudorapidity and azimuthal projections for all multiplicity intervals and event-shape classes.
This implies that the transverse-momentum correlations quantified by $P_2^{CD}$ are confined to a narrower range of relative coordinates than the balancing-charge correlations quantified by $B$. 

The differences in the widths of the charge balance function and
the charge-dependent transverse-momentum balance function can be
understood in terms of the different sensitivities of these
observables to particle production. The charge balance function, $B$, characterizes the
relative distribution of balancing charges, whereas $P_2^{\rm CD}$
introduces an additional constraint by weighting particle pairs
according to their transverse-momentum correlations. Consequently, $P_2^{\rm CD}$ is sensitive not only to the relative separation of particle pairs but also to their correlated transverse-momentum. Particle production mechanisms such as jet fragmentation and resonance decays can contribute to these correlations and influence the shape of $P_2^{\rm CD}$ \cite{Behera:2025ymi}. This difference in sensitivity may help explain the narrower widths and sharper near-side structure observed for $P_2^{\rm CD}$ compared with $B$.

\section{Summary and conclusions}
\label{summary}
This paper reports the study of charge balance function $B$ and the charge-dependent transverse momentum balance function $P_{2}^{\rm CD}$ in $p + p$ collisions at $\sqrt{s}=13.6$~TeV using \texttt{PYTHIA~8.315}, with flattenicity as the event shape observable. Both the studied observables show strong near-side correlations whose strength increases with increase in charged particle multiplicity. 
The RMS widths $\sigma_{B}$ and $\sigma_{P_{2}^{CD}}$ decrease with increasing multiplicity, consistent with previous studies. The dependence of the balance functions on flattenicity further demonstrates that the structure of the two-particle correlations is sensitive to event topology. Jet-like events exhibit stronger and narrower correlations than isotropic events,  possibly due to the
localized particle production associated with jet fragmentation. The stronger flattenicity dependence of $P_{2}^{\rm CD}$ indicates that it is more sensitive to
event topology than $B$, which primarily probes charge-dependent particle-pair correlations. Together, these observables provide complementary information on charge balancing and transverse-momentum correlations, offering a more differential perspective on particle production in high-energy $p+p$ collisions.

\appendix

\acknowledgments

S.D.R. acknowledges financial support from the Department of Science and Technology (DST), Government of India, through the DST-INSPIRE Fellowship (IF240391).





\begingroup\raggedright\endgroup

\end{document}